\documentclass[10pt,conference]{IEEEtran}
\IEEEoverridecommandlockouts

\usepackage{cite}
\usepackage{amsmath,amssymb,amsfonts}
\usepackage{algorithmic}
\usepackage{graphicx}
\usepackage{textcomp}
\usepackage{xcolor}
\usepackage{tikz}
\usepackage{color}
\usepackage{listings}
\usepackage{xspace}
\usepackage{pifont}
\usepackage{tcolorbox}
\usepackage{url}
\usepackage{subcaption}

\newcommand*{\ie}{i.e.,\@\xspace}
\newcommand*{\eg}{e.g.,\@\xspace}

\newcommand{\revision}[1]{\textcolor{black}{#1}}

\newcommand*\circled[1]{\tikz[baseline=(char.base)]{
            \node[shape=circle,draw,inner sep=2pt] (char) {#1};}}

\newcommand{\rqone}{\textit{To what extent does the usage of custom tags in prompts improve the energy efficiency of Llama 3 while performing code completion tasks?}}

\newcommand{\rqtwo}{\textit{How do custom tags influence predictive accuracy of Llama 3 while performing code completion tasks?}}

\newcommand{\os}{\textit{one-shot}\xspace}
\newcommand{\zs}{\textit{zero-shot}\xspace}
\newcommand{\fs}{\textit{few-shots}\xspace}

\newboolean{showcomments}

\setboolean{showcomments}{true}

\ifthenelse{\boolean{showcomments}}
  {
}

\newcommand*{\etal}{\emph{et~al.}\@\xspace}

\definecolor{eclipseStrings}{RGB}{42,0.0,255}
\definecolor{eclipseKeywords}{RGB}{127,0,85}
\colorlet{numb}{magenta!60!black}

\lstdefinelanguage{json}{
	basicstyle=\small\ttfamily,
	commentstyle=\color{eclipseStrings}, 
	stringstyle=\color{eclipseKeywords}, 
	numbers=left,
	numberstyle=\scriptsize,
	stepnumber=1,
	numbersep=6pt,
	showstringspaces=false,
	breaklines=true,
	frame=lines,
	string=[s]{"}{"},
	comment=[l]{:\ "},
	morecomment=[l]{:"},
	literate=
	*{0}{{{\color{numb}0}}}{1}
	{1}{{{\color{numb}1}}}{1}
	{2}{{{\color{numb}2}}}{1}
	{3}{{{\color{numb}3}}}{1}
	{4}{{{\color{numb}4}}}{1}
	{5}{{{\color{numb}5}}}{1}
	{6}{{{\color{numb}6}}}{1}
	{7}{{{\color{numb}7}}}{1}
	{8}{{{\color{numb}8}}}{1}
	{9}{{{\color{numb}9}}}{1}
}

\def\BibTeX{{\rm B\kern-.05em{\sc i\kern-.025em b}\kern-.08em
    T\kern-.1667em\lower.7ex\hbox{E}\kern-.125emX}}

\begin{document}

\title{Prompt engineering and its implications on the energy consumption of Large Language Models }

\author{
	\IEEEauthorblockN{Riccardo Rubei}
	\IEEEauthorblockA{
		\textit{University of L'Aquila}\\
		L'Aquila, Italy  \\
		riccardo.rubei@univaq.it}
\and
	\IEEEauthorblockN{Aicha Moussaid}
	\IEEEauthorblockA{
		\textit{University of L'Aquila}\\
		L'Aquila, Italy  \\
		aicha.moussaid@student.univaq.it}
\and
	\IEEEauthorblockN{Claudio Di Sipio}
	\IEEEauthorblockA{
		\textit{University of L'Aquila}\\
		L'Aquila, Italy \\
		claudio.disipio@univaq.it}
\and
	\IEEEauthorblockN{Davide Di Ruscio}
	\IEEEauthorblockA{
		\textit{University of L'Aquila}\\
		L'Aquila, Italy \\
		davide.diruscio@univaq.it}
}

\maketitle

%
%
%
%

\begin{abstract}
    Reducing the environmental impact of AI-based software systems has become critical. The intensive use of large language models (LLMs) in software engineering poses severe challenges regarding computational resources, data centers, and carbon emissions. In this paper, we investigate how prompt engineering techniques (PETs) can impact the carbon emission of the Llama 3 model for the code generation task. We experimented with the CodeXGLUE benchmark to evaluate both energy consumption and the accuracy of the generated code using an isolated testing environment.   
    Our initial results show that the energy consumption of LLMs can be reduced by using specific tags that distinguish different prompt parts. Even though a more in-depth evaluation is needed to confirm our findings, this work suggests that prompt engineering can reduce LLMs' energy consumption during the inference phase without compromising performance, paving the way for further investigations.
\end{abstract}
     
    

    \section{Introduction}
    \label{sec:intro}
    The environmental impact of software systems has been a growing concern in recent years \cite{verdecchia2021green,georgiou2017analyzing}, thus fostering the development of green software engineering (GSE) \cite{calero_green_2015} by proposing dedicated methodologies \cite{GULDNER2024402}, frameworks \cite{pyRAPL,noureddine-ie-2022}, and guidelines \cite{MANCEBO2021100558}. Nevertheless, the rise of AI-intensive systems has posed new challenges regarding energy consumption and carbon emissions \cite{Strubell_Ganesh_McCallum_2020}.

In particular, both training and querying large language models (LLMs) to outperform traditional techniques in code-related tasks \cite{tufano_using_2022,mastropaolo_studying_2021,wang_bridging_2022} is computationally expensive and requires large amounts of resources and has a significant carbon footprint \cite{castano_exploring_2023}. Moreover, assessing them is challenging due to \textit{i)} higher variability in the generated code and \textit{ii)} the lack of standardized guidelines and information for measuring carbon emissions even in dedicated model repositories \cite{castano_exploring_2023}. While a plethora of approaches have been proposed to measure the impact on the hardware \cite{SamsiZMLMJBKTG23}, we focus on the usage of prompt engineering techniques (PETs) to mitigate the energy consumption of LLMs during the inference phase while supporting the code completion task. By relying on the CodeXGLUE \cite{lu1codexglue} dataset, we first devise a dedicated component that selects and tests different prompts on Llama 3 \cite{dubey2024llama3herdmodels} to assess their impact on the energy consumption using the CodeCarbon tool \cite{codecarbon}. Concretely, we used traditional PETs as baselines and devise four additional configurations using additional tags and explanations to enhance the baseline prompts. 
In particular, we aim to answer the following research questions:

\noindent \ding{226} \textbf{RQ$_1$}: {\rqone} We explore the effects of specifically introduced custom tags on the energy consumption of LLMs during the inference phase to support code completion tasks. To this end, we first calculate the energy consumption of three well-known PETs, \ie \zs, \os, and \fs, without any modifications. Afterward, we compare these baseline prompts with an enhanced version using additional tags that we introduced to help the inference phase of the model. In addition, we also measure the overall time required to perform the assigned task.

\noindent \ding{226} \textbf{RQ$_2$}: {\rqtwo} We aim to analyze the impact of custom tags on the performance of Llama 3, focusing on well-established accuracy metrics,  \ie exact matches and edit distance. We chose to use these metrics because they have been successfully applied in the CodeXGLUE benchmark and are recognized as effective tools for evaluating code completion when using LLMs \cite{HuseinAC25}.

\begin{figure*}[t!]
	\centering
	\includegraphics[width=\linewidth]{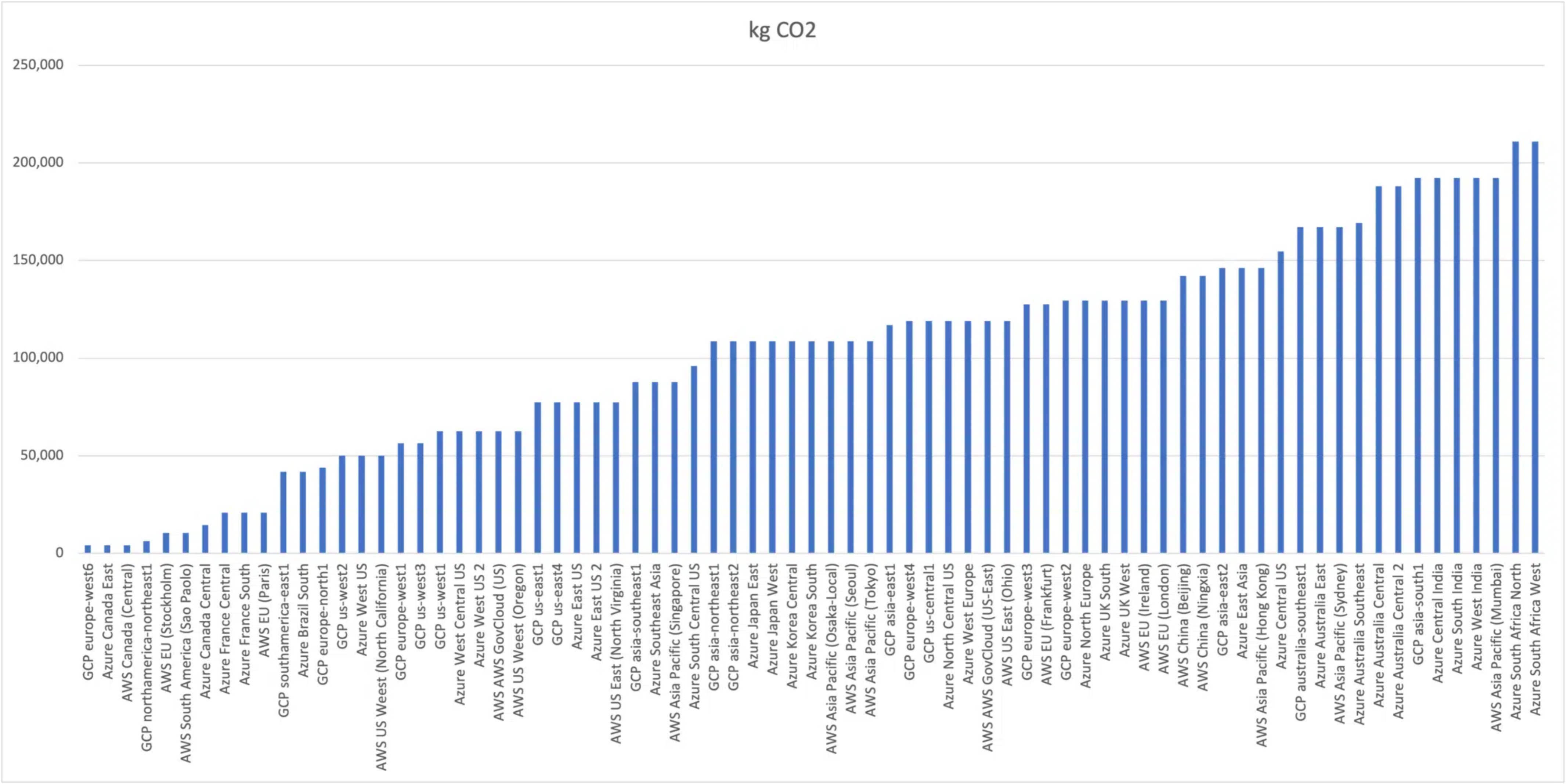}
	\caption{Carbon emissions of GPT-3 models as reported in \cite{shrinkthatfootprint2023}.}
	\label{fig:motivation}
\end{figure*}

Our findings reveal that the energy consumption of LLMs for the inference phase can be reduced by using the introduced custom tags. Moreover, we show that the energy consumption of LLMs is highly dependent on the used PETs. Although further experimentation involving additional tasks and LLMs is needed, the presented work suggests that prompt engineering can play a key role in reducing the energy consumption of LLMs without compromising their performance.

The main contributions of this work are as follows:
\begin{itemize}
\item We investigate the effects of several prompt engineering techniques and custom tags on the energy consumption of LLMs while performing code completion tasks;
\item Our research examines the trade-offs between energy consumption in terms of carbon emission, execution time, and generated code accuracy to investigate the balance between energy efficiency and model accuracy;
\item We provide a replication package\footnote{\url{https://github.com/riccardoRubei/Greens-2025-Replication-Package}} to foster further research on the topic.
\end{itemize}

    \section{Background}
    \label{sec:background}
    

While measuring traditional software impact in terms of emissions is well-established \cite{verdecchia2021green,MANCEBO2021100558}, assessing LLMs consumption is still challenging, as High-Performance Computing (HPC) clusters are often required to run the training process, which can last for weeks or even months. Therefore, measuring the energy consumption in terms of carbon emissions is particularly challenging in those environments due to several factors, \eg parallel jobs or the non-exclusive use of the cluster.

Moreover, even well-maintained LLMs leaderboard benchmarks \cite{trustbit_llm_benchmarks,lmarena_leaderboard,oobabooga_benchmark} do not report energy consumption, focusing instead on accuracy metrics. Figure \ref{fig:motivation} shows the carbon emissions of the GPT-3 model in different server regions for three big IT players, \ie Google, Amazon, and Microsoft. For instance, some models emit carbon that is equivalent to the average of five cars over their lifetimes \cite{strubell2019energypolicyconsiderationsdeep}, thus underlining significant sustainability concerns, especially when considering the growing scope of LLM-based implementations and their integration into everyday life.  This highlights the need to reduce the carbon footprint of LLMs and to examine the details that contribute to the reported figures.

To address the environmental impact of software, a range of energy monitoring tools \cite{pyRAPL,noureddine-ie-2022} has been recently developed to measure the carbon emissions associated with code execution. Among these, the CodeCarbon tool \cite{codecarbon} is a widely adopted Python library that estimates the energy consumption of code executions. It can also calculate the carbon footprint by measuring the electricity power consumption of the underlying hardware architecture, \ie GPU, CPU, and RAM. In addition, it can estimate the carbon intensity of the region where the computing is done. This study focuses on the energy consumption related to GPU usage without considering the carbon emission. 

Concerning the inference phase of LLMs, prompt engineering is pivotal to enhancing LLMs' generation capabilities. The most basic PET is \zs, in which the LLM is given a query without any example of outputs, which are expected from the given inputs~\cite{10.5555/3045118.3045347}. In contrast, \os prompting provides the model with a single example, offering a minimal context to guide responses. The \fs prompting \cite{few-shot} involves multiple examples, allowing the model to generalize more effectively with limited supervision~\cite{LI2024112002}. In the scope of this paper, we focus on different \textit{shot} techniques \ie \zs, \os, and \fs given their efficiency and success in improving the performance of LLMs in source code-related tasks.

Quantization \cite{gholami2021surveyquantizationmethodsefficient} is a technique that reduces the computational and memory requirements of LLMs by lowering the precision of their numerical representations (\eg from 32-bit to 8-bit). This compression speeds up inference, making LLMs more efficient with minimal impact on accuracy. In this paper, we leverage quantization alongside PETs to minimize the computational cost while maintaining performance in code-related tasks.

\begin{figure*}[t!]
	\centering
	\includegraphics[width=.8\textwidth,keepaspectratio]{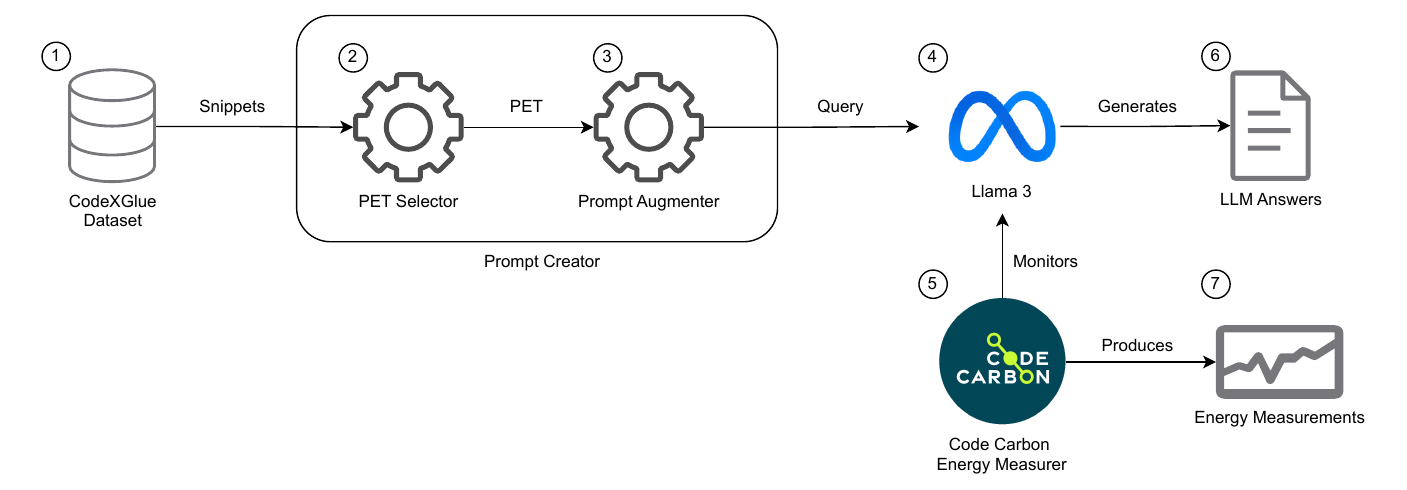}
	\caption{Workflow of the performed experiments.}
	\label{fig:executionProcedure}
\end{figure*}

While developing a comprehensive methodology for measuring LLM energy consumption is beyond this paper's scope, we focus on reducing these emissions through efficient PETs. By utilizing custom tags, we aim to lower energy consumption in LLMs used for code-related tasks, offering an approach that balances sustainability with performance.
       
    \section{Performed experiments}
    \label{sec:methodology}


%

Figure \ref{fig:executionProcedure} depicts the workflow of the experiments we performed to answer the two research questions. Starting from the CodeXGLUE dataset \cite{lu1codexglue} {\small \circled{1}}, \textit{prompt creator} {\small \circled{2}} translates input prompts into a format that Llama 3 can understand, before augmenting them with tags that we specifically introduced {\small \circled{3}}. 
%
%
Afterward, the crafted prompts are used to query the LLM locally deployed {\small \circled{4}}. For each snippet, we executed 75 queries.\footnote{Three prompting techniques (\ie \zs, \os, and \fs) $\times$ five prompt configurations $\times$ five repetitions to mitigate possible energy measurement inaccuracies.} Each Llama run is monitored {\small \circled{5}} by the CodeCarbon energy monitoring tool. 
For each execution, we store three artifacts (question, answer {\small \circled{6}}, and measurement {\small \circled{7}}), to enable both efficiency and accuracy analysis.


\subsection{Dataset}


 Among different benchmarks, we select CodeXGLUE as it is tailored for supporting and evaluating LLMs in code-related tasks \cite{faiz2023llmcarbon,10.1145/3540250.3549113}. In this paper, we consider the code completion task as it is widely supported by LLMs as recently investigated \cite{10.1145/3695988,10.1145/3661167.3661215}. This task leverages established evaluation methodologies in the literature, enabling straightforward comparisons with ground truth data.
\subsection{Prompt Creator}\label{sec:prompt_creator}
This component is responsible for defining and augmenting prompts that have been used to query the model under analysis. In particular, we use standard PETs, \ie \zs, \os, and \fs, as a baseline to evaluate the effect of custom tags in terms of energy impact.
The Llama 3 model card\footnote{www.llama.com/docs/model-cards-and-prompt-formats/meta-llama-3/} defines several tokens which form the model's input. 
We aim to investigate the impact of custom tags on energy consumption and performance metrics for Llama 3. To this end, we define five distinct prompt configurations. Each prompt comprises two key components: a \texttt{role} attribute and \texttt{content} specification, as illustrated in Listing \ref{lst:Zero-Shot-Default}. The \texttt{role} attribute can be assigned as either \texttt{system} or \texttt{user}. In the case of \texttt{system}, the accompanying \texttt{content} attribute specifies the task to be performed, thus clarifying the expected contribution from the model. For example, in Listing \ref{lst:Zero-Shot-Default}, the system role is configured to instruct the model on a code completion task for given code fragments. The \texttt{user} role, on the other hand, introduces the input code snippet that the model is expected to complete.

%
%
%
%
%

\revision{According to the different configurations, the \texttt{content} can be enhanced with custom tags or explanations related to the task. The configurations are defined as follows:}


\medskip
\noindent
\textit{$C_0$ - default}: We define the model's role and provide the incomplete snippet without any customization. In the case of \os and \fs, we describe one and five examples, respectively. We fix the number of examples equal to five for the \fs technique since it obtains adequate accuracy with limited token size \cite{10.1145/3551349.3559555}. Nonetheless, we acknowledge that a deep study concerning the different shot sizes is needed. Listing \ref{lst:Zero-Shot-Default} depicts an example of prompt in its default representation. 

\begin{lstlisting}[basicstyle=\small,language=json,frame=single,numbers=right, caption={Example of a \zs prompt.}, label={lst:Zero-Shot-Default}]
{
"role": "system",
"content" : "You are an AI assistant specialized in code completion for Java. Your task is to complete the provided Java code segment with one line. Give only the code completion.",
},{
"role": "user",
"content": "package com.lmax.disruptor.support; import java.util.concurrent.ThreadFactory; public final"
}
\end{lstlisting}

\medskip
\noindent
\textit{$C_1$ - use of custom tags without explanation}:  We augment prompts by using custom tags \ie $<$code$>$ and $<$incomplete$>$ to support the inference phase to distinguish the input source code, and the fragment that needs to be completed. We do not provide any explanation of what is the meaning of such custom tags. Therefore, we aim to explore the LLM's capability to understand the customization. Listing \ref{lst:Zero-Shot-Code-Customized} is an example of a code fragment augmented with custom tags. 
\begin{lstlisting}[basicstyle=\small,language=json,frame=single, caption={Fragment of a prompt including custom tags.},label={lst:Zero-Shot-Code-Customized}]
{	
"role": "user",
"content" :"<code> package com.lmax.disruptor.support;  import java.util.concurrent.ThreadFactory; </code> <incomplete> public final </incomplete>"
}
\end{lstlisting}

\medskip
\noindent
\textit{$C_2$ - use of custom tags with explanation}: We embed the meaning of the custom tags in the prompt as shown in Listing~\ref{lst:Zero-Shot-Code-Customized-Explained}.

\begin{lstlisting}[basicstyle=\small,language=json,frame=single, caption={Fragment of a prompt including custom tags explanation.},label={lst:Zero-Shot-Code-Customized-Explained}]
{	
"role": "user",
"content" :"The code to analyze is marked by the <code> tag and the line to be completed is marked by the <incomplete> tag. <code> package com.lmax.disruptor.support;  import java.util.concurrent.ThreadFactory;</code><incomplete> public final </incomplete>"
}
\end{lstlisting}

\medskip
\noindent
\textit{$C_3$ - custom prompt explained in the system}: Differently from configuration $C_2$, the explanation of custom tags is given in the system role part of the input prompt as shown in Listing \ref{lst:Zero-Shot-Customized}.

\begin{lstlisting}[basicstyle=\small, language=json, frame=single, caption={Example of a \zs prompt including the definition of custom tags.}, label={lst:Zero-Shot-Customized}]
{
"role": "system",
"content" : "You are an AI assistant specialized in code completion for Java. Your task is to complete the provided Java code segment with one line. Give only the code completion. The code to analyze is marked by the <code> tag and the line to be completed is marked by the <incomplete> tag.",
},{
"role": "user",
"content": "<code> package com.lmax.disruptor.support;  import java.util.concurrent.ThreadFactory;</code><incomplete> public final </incomplete>"
}
\end{lstlisting}

\medskip
\noindent
\textit{$C_4$ - no system definition}: With this configuration, we want to assess the effect of the complete absence of the system role definition. Therefore, we provide only the incomplete input snippet and a task definition directly in the prompt without any customization as illustrated in Listing \ref{lst:Zero-Shot-No-Role}.


\begin{lstlisting}[basicstyle=\small,language=json,frame=single, numbers=right, caption={Fragment of a prompt including custom tags.},label={lst:Zero-Shot-No-Role}]
{
"role": "system",
"content" : "",
},{	
"role": "user",
"content" :"Hi, complete the following snippet adding one line please: package com.lmax.disruptor.support;  import java.util.concurrent.ThreadFactory; public final"
}
\end{lstlisting}


The process ends with the generation of three different artifacts, \ie questions, answers, and measurements. A question is a copy of the query given to the LLM and it is stored for subsequent analysis. The measurement is the outcome of the Llama 3 process monitored by CodeCarbon to solve the code completion task. Meanwhile, an answer is just a sequence of Java statements to complete the input snippet. In some cases, the LLM answer is verbose. Therefore, we can notice a sequence of several lines of code.
\subsection{Metrics}
Concerning the metrics, we rely on CodeCarbon predefined format\footnote{https://mlco2.github.io/codecarbon/output.html} to avoid any bias in the comparison. Since our study focuses on the energy effects on the GPU, we rely on the \textit{gpu\_energy} value to support the evaluation.

During our investigation, we evaluate the effects of prompting techniques and customization of the prompts. Therefore, we employ the following metrics:

\noindent \ding{228} \textbf{Energy Consumption:} This metric quantifies the energy consumed during the inference phase of Llama excluding the model loading. We rely on the calculation provided by CodeCarbon. In its report, we focus on the value of \textit{gpu\_energy} which calculates the energy consumed during in the inference, expressed in kWh. To reduce biases related to unprecise monitoring, we repeated the tests 5 times, calculating eventually the average.

\noindent \ding{228} \textbf{Execution Time:} The execution time calculates the duration needed by Llama 3 to perform the inference. The monitoring is limited only on the inference phase, excluding the model loading time. The time is excerpted from the CodeCarbon report similarly for the energy value.

\noindent \ding{228} \textbf{Edit Distance:} The edit distance metric calculates how similar the proposed answer is to the ground truth, by counting the  number of characters that need to be substituted, inserted, or deleted to transform an input string into a target one. We used the nltk edit distance, which implements the well-known Levenshtein Distance \cite{levenshtein}. 
%


\noindent \ding{228} \textbf{Exact Match} The exact match metric measures whether the answer of the LLM has an edit distance of 0, meaning that the ground truth and answer are the same. Since LLMs are generally verbose, we fixed the exact match threshold to edit distance less or equal to 2. The rationale is that Llama produces the results by adding several random characters to the answer, e.g. extra spaces, single and double quotes, semicolons. 


\subsection{Execution process}

The experiments have been performed by considering the settings shown in Table \ref{tab:configuration}. 
%
%
In particular, we tested 1,000 random incomplete Java snippets retrieved from the code-completion dataset of CodeXGLUE. As discussed in Section \ref{sec:results} the overall execution requires more than 250 hours. We calculated an average test time per snippet of about 900 seconds. Therefore, we limit ourselves to 1,000 snippets. with the abovementioned PETs 

\begin{table}[t!]
	\centering
	\caption{Summary of the Experimental Settings}
	\label{tab:configuration}
	\small
	\begin{tabular}{|l|l|}
		\hline
		\textbf{Model} & Llama3 8B - Instruct \\ \hline
		\textbf{Snippets} & 1,000 \\ \hline
		\textbf{PETs} & 3 \\ \hline
		\textbf{Custom Prompts} & 5 \\ \hline
		\textbf{Repetitions} & 5 \\ \hline
		\textbf{Pause} & 10 seconds \\ \hline
		\textbf{Metrics (Performance)} & Energy Consumption, Execution Time \\ \hline
		\textbf{Metrics (Accuracy)} & Exact Match, Edit Distance \\ \hline
	\end{tabular}
\end{table}

As discussed in Section \ref{sec:prompt_creator}, we defined five distinct configurations for each query. Consequently, we tested every combination of prompting techniques and the use of custom tags. To ensure experimental reliability, each test is repeated five times \cite{10.1145/3510003.3510221,10174114}, with a ten-second pause between each test to mitigate potential tail effects \cite{10174114,bornholt2012model}. We use two metrics to evaluate energy consumption and execution time, and two primary metrics (exact match and edit distance) to assess the impact of different configurations on accuracy. These metrics align with those used in the original evaluation of the code completion benchmark suite by the authors of CodeXGLUE.

All the experiments have been conducted on an isolated desktop equipped with an AMD Ryzen 7 5800X 3.8GHz CPU and an Nvidia Geforce RTX 4060 TI (8 GB VRAM).\footnote{https://www.nvidia.com/en-us/geforce/graphics-cards/40-series/rtx-4060-4060ti/} The operating system is Xubuntu 23.04.
Since the GPU provided only 8GB of RAM, we used the quantized version of the Llama model \ie we used 16-bit float rather than the default 32-bit.


    \section{Experimental Results}
    \label{sec:results}

\begin{figure*}[t!]
	\centering
	\begin{subfigure}[b]{0.5\textwidth}
		\centering
		\includegraphics[width=0.9\textwidth,keepaspectratio]{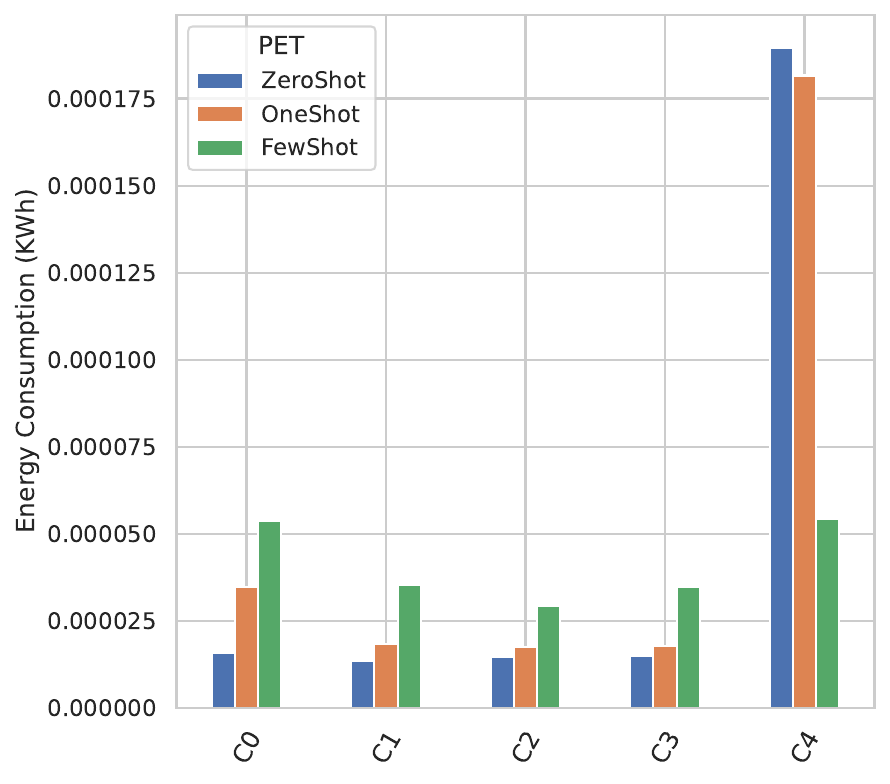}
		\caption{Energy Consumption in (kWh).}
		\label{fig:RQ2_Energy_Comparison}
	\end{subfigure}
	\hfill
	\begin{subfigure}[b]{0.45\textwidth}
		\centering
		\includegraphics[width=0.9\textwidth,keepaspectratio]{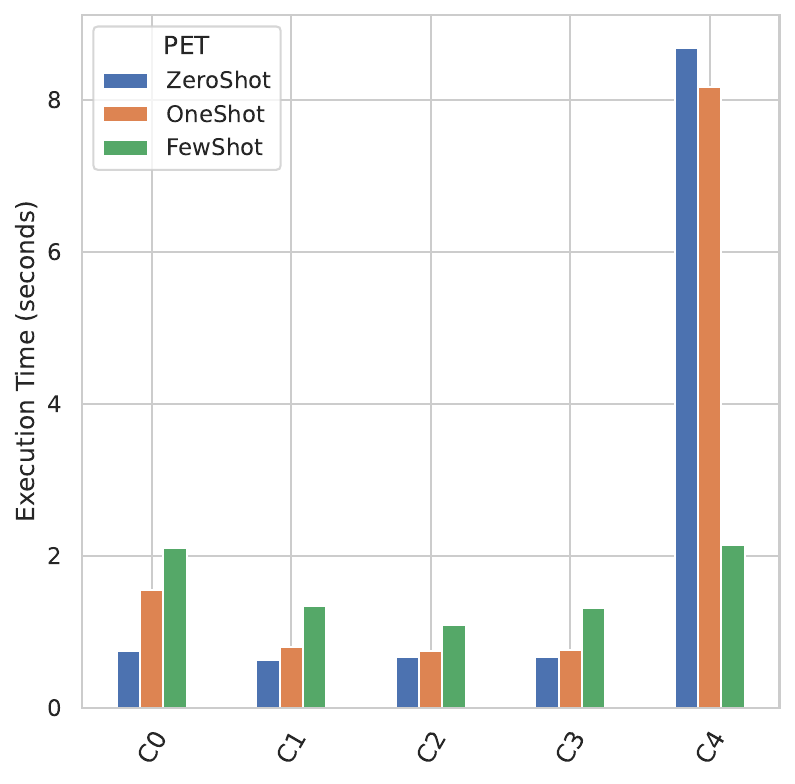}
		\caption{Execution Time.}
		\label{fig:RQ3_Execution_Time}
	\end{subfigure}
	\caption{Energy consumption with different prompt configurations.}
	\label{fig:energy}
\end{figure*}

\begin{figure*}[t!]
	\centering
	\begin{subfigure}[b]{0.45\textwidth}
		\centering
		\includegraphics[width=0.9\textwidth,keepaspectratio]{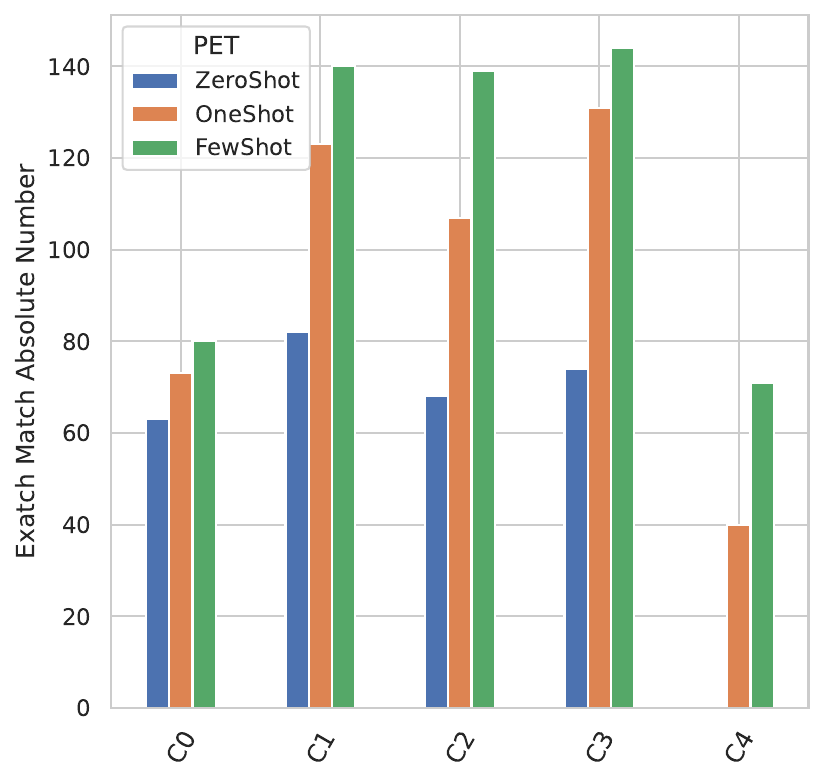}	
		\caption{Exact Match.}
		\label{fig:RQ3_Exact_Match}
		
	\end{subfigure}
	\hfill
	\begin{subfigure}[b]{0.45\textwidth}
		\centering
		\includegraphics[width=0.9\textwidth,keepaspectratio]{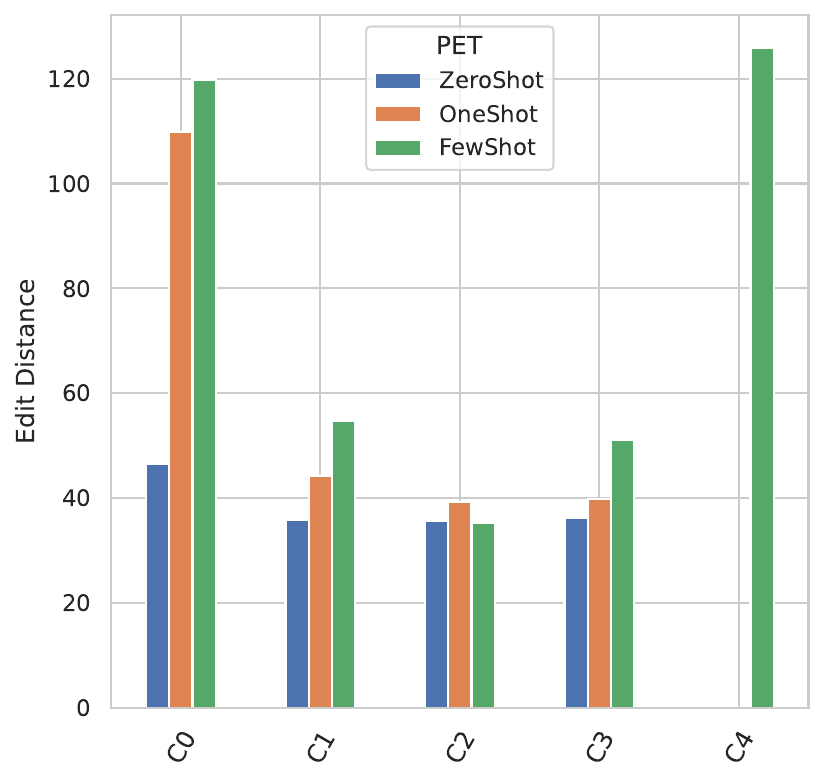}		
		\caption{Edit Distance.}
		\label{fig:RQ3_Edit_Distance}		
	\end{subfigure}
	\caption{LLMs accuracy with different prompt configurations.}
	\label{fig:acc} 
\end{figure*}

\subsubsection*{Answering \textbf{RQ$_1$}} 
Figure \ref{fig:RQ2_Energy_Comparison} shows the energy consumption of the three prompt techniques applied to the five different configurations. In particular,  with the default configuration \textit{C0}, \zs is the most energy-efficient, with an average cost of about 0.000016 kWh. \os and a \fs consumed an average of 0.000035 kWh and 0.000054 kWh, respectively. 

Custom tags can contribute to reducing the energy consumption of the video card. As shown in Fig. \ref{fig:RQ2_Energy_Comparison}, the best configuration is the \textit{C2} (explanation in prompts). While the \zs technique passed from 0.0000157 (of \textit{C0}) to 0.0000146 (-7\%), \os and \fs reduced the consumption from 0.0000347 to 0.0000174 (-99\%) and from 0.0000537 to 0.0000293 (-83\%) comparing with the default configuration \textit{C0}, respectively. It is also interesting to see the results of \textit{C4}, in which we do not specify any role in the system token. The consumption increased from 0.0000157 to 0.000189 kWh for \zs and from 0.0000347 to 0.000181 kWh for \os.  
The reason is that the model started to generate completely new code snippets when asked to finalize the code given as input. 
The \fs technique seems to be less affected by this problem. The sequence of example questions and answers instructed the model on the behaviour despite the lack of the system role specification. 

Concerning the execution time, Figure \ref{fig:RQ3_Execution_Time} reports the results obtained for all the prompt configurations. Similar to energy consumption, the usage of custom tags provides a general improvement in performance. In particular, the \os and \fs reduced the average time from 1.54 seconds of configuration \textit{C0} to 0.74 (-52\%) and from 2.1 to 1.09 (-48\%), respectively, using configuration \textit{C2}. The \zs technique performed better using  \textit{C1}, reporting an improvement from 0.74 seconds to 0.63 (-14.8\%). Similarly, for the energy consumption, in the case of \textit{C4}, we can notice a remarkable increase in execution time for \zs and \os. 

\begin{tcolorbox}[colframe=gray!80]
\textbf{Answer to RQ$_1$:} Our study reveals that custom tags can reduce the energy consumption of LLMs across the three prompt engineering techniques tested for source code completion tasks. 
\end{tcolorbox}

\subsubsection*{Answering \textbf{RQ$_2$}} Figure \ref{fig:acc} depicts the obtained results in terms of accuracy metrics. In particular, Figure \ref{fig:RQ3_Exact_Match} shows the effects of custom tags on exact match performance across different prompt engineering techniques. Overall, we observe an increase in exact matches for configuration C1-C3 in comparison with the default configuration C0. Notably, \zs shows the greatest improvement with \textit{C1}, where exact matches rise from 63 to 82, reflecting a 23\% increase. Both \os and \fs see substantial gains with \textit{C3}, achieving approximately a 44\% improvement. Interestingly, with \textit{C4}, \zs fails to achieve any exact matches.
 
Figure \ref{fig:RQ3_Edit_Distance} shows the impact of custom tags on edit distance metrics, where an edit distance of 0 indicates a perfect result.  Overall, custom tags contributed to a reduction in edit distance, with \textit{C2} emerging as the most effective configuration across all prompt engineering techniques. Specifically, \zs showed a 24\% improvement, \os achieved a 64\% reduction, and \fs improved by 70\%.
Results for \zs and \os are omitted for \textit{C4} because, with this configuration, the LLM produced uncontrolled responses. As a result, it was impossible to calculate edit distance accurately, as the outputs included both code and explanatory text.
Despite lacking explicit role definitions, \fs continued to yield satisfactory results.

\begin{tcolorbox}[colframe=gray!80]
	\textbf{Answer to RQ$_2$:} Prompt customizations enhanced the accuracy of the tested PETs, showing a positive trend with increased exact matches and reduced edit distances. 
	\end{tcolorbox}  

    \section{Related work}
    \label{sec:relatedwork}
    
\subsubsection*{Assessing LLMs energy consumption}

Jagannadharao  \etal \cite{jagannadharao_timeshifting_2023} investigate the usage of time-shifting technique to reduce the energy consumption of LLMs during long-running training sessions. Concretely, the authors estimates the consumption of Llama model by pausing and resuming the training when the carbon emission is below a certain threshold. The results shows that the proposed approach succeed in reducing the carbon emission even though the region may impact the obtained results.
Liu and Yin \cite{liu_green_2024} investigate how to reduce and measure the consumption of pre-trained models by combining fine-tuning and efficient tokenizers. In particular, BERT, DistilBERT, and T5 models are compared using SQuAD benchmark \cite{rajpurkar2016squad} in terms of accuracy and carbon emissions. The experimental results reveal that both the T5 and BERT models emitted considerably more CO2 compared to DistilBERT and the T4 GPU contributes in reducing the overall carbon emissions. Samsi \etal \cite{SamsiZMLMJBKTG23} compare the inference performance in terms of watts of different Llama models, \ie evaluating smaller models (7B, 13B) against the largest available version (65B) at the time of writing. In addition, the authors consider different GPUs, \ie V100 and A100. The study reveals that 8 V100 GPUs each with 32 GB of RAM or 4 A100 GPUs each with 80GB of memory are required for any meaningful inferences with the 65B LLaMA model, thus making small models a suitable choice for energy-efficient applications. 
Cursaro \etal \cite{10.1007/978-3-031-70245-7_12} conduct a controlled experiment in which code generated by CodeLlama is compared with the human one considering different languages, \ie C++, Java, and Python, tested on a dedicated platform. The results show that explicitly asking  to generate energy-efficient code results in an equal or worse energy efficiency. In our work, we focus on reducing energy consumption of Llama by customizing the prompt and using a dedicated GPU.

\subsubsection*{Prompt customization} Fagadau \etal \cite{10.1145/3643916.3644409} explored the influence of eight prompt features on Copilot's code outputs, analyzing 124,800 prompts designed to implement 200 Java methods. The findings indicate that prompts including concise method summaries and examples lead to higher accuracy in generated code while additional details as boundary cases have a negative impact. 
Reynolds and McDonell \cite{10.1145/3411763.3451760} explored example-free strategies in prompt engineering, aiming to enhance results by refining prompt structure. In particular, they embody analogies and synonyms during task specification and limit undesired outputs with negative prompting. Li \etal \cite{li2023cctesttestingrepairingcode} investigate prompt modifications using metamorphic testing. Using Copilot as baseline model, code fragments are injected in the prompts instead of natural language. Then, semantic mutations are introducted to modify the prompts. Similar to our approach, Wang \etal \cite{10.1145/3540250.3549113} proposes prompt tuning, a novel PET executed during the fine-tuning process. This technique involves the \textit{soft prompting} in which task-related knowledge are tagged using virtual tokens instead of using fixed annotation, \ie \textit{hard prompting}. The empirical evaluation conducted on CodeBERT and CodeT5 shows that prompt tuning consistently outperforms fine-tuning in three code-related tasks, \ie defect prediction, code summarization, and code translation. Compared to those works, we introduce explanations in prompts to reduce the energy consumption of Llama 3 model in code generation task. 

    \section{Threats to validity}
    \label{sec:threats}
    This section discusses threats that may hamper the results of our study and corresponding mitigation strategies.

\textit{Internal validity} concerns factors that may impact the measurements, \ie noise interference, background processes, and voltage fluctuations. To mitigate these issues, all the experiments have been conducted in an isolated Linux-based system without parallel or background tasks running on the GPU. In addition, we repeated each experiment five times and a 10-second pause between each query execution to prevent potential performance degradation and statistical anomalies, thus increasing the reliability of  measurements.


Threats to \textit{external validity} are related to the generalizability of the performed experiments, \ie the obtained results in terms of energy consumption and accuracy may vary considering different tasks and LLMs. Concerning the data, we employed CodeXGLUE, a well-known dataset exploited in several studies. We were forced to cap our dataset to 1,000 snippets, since the time needed to test one snippet has been evaluated to 900 seconds. Finally, the measurements calculated on the inference without any customization are strictly related to the particular task that we decided to study, thus code generation or text summarization might require different energy resources. We mitigated this threat focusing on the effects of the customization. 

    \section{Conclusion and future work}
    \label{sec:conclusion}
    Motivated by the increasing carbon emissions of LLMs, we proposed a preliminary investigation on the effects of prompt customizations on Llama 3 model for the specific task of code completion. Our results show that augmenting prompts with dedicated custom tags and explanations succeed in reducing the energy consumption yet preserving adequate accuracy. In particular, with the best configuration, \zs reduced the consumption of about 7\%, whereas \os and \fs decreased their consumption of about 99\% and 83\%, respectively. For future work, we plan to extend the study to additional LLMs and code-related tasks. In addition, we will investigate advanced techniques, \eg retrieval augmented generation (RAG) or fine-tuning, to further reduce the carbon emissions of LLMs. Finally, we plan to investigate the effects of custom prompts in different software engineering tasks.

    \section*{Acknowledgments}

This work has been partially supported by the EMELIOT national research project, which has been funded by the MUR under the PRIN 2020 program grant n. 2020W3A5FY, the European Union--NextGenerationEU through the Italian Ministry of University and Research, Projects PRIN 2022 PNRR \emph{``FRINGE: context-aware FaiRness engineerING in complex software systEms''} grant n. P2022553SL, and the Italian ``PRIN 2022'' project \emph{``TRex-SE: Trustworthy Recommenders for Software Engineers,''} grant n. 2022LKJWHC.

\bibliographystyle{IEEEtran}
\bibliography{main}

\begin{thebibliography}{10}
\providecommand{\url}[1]{#1}
\csname url@samestyle\endcsname
\providecommand{\newblock}{\relax}
\providecommand{\bibinfo}[2]{#2}
\providecommand{\BIBentrySTDinterwordspacing}{\spaceskip=0pt\relax}
\providecommand{\BIBentryALTinterwordstretchfactor}{4}
\providecommand{\BIBentryALTinterwordspacing}{\spaceskip=\fontdimen2\font plus
\BIBentryALTinterwordstretchfactor\fontdimen3\font minus \fontdimen4\font\relax}
\providecommand{\BIBforeignlanguage}[2]{{%
\expandafter\ifx\csname l@#1\endcsname\relax
\typeout{** WARNING: IEEEtran.bst: No hyphenation pattern has been}%
\typeout{** loaded for the language `#1'. Using the pattern for}%
\typeout{** the default language instead.}%
\else
\language=\csname l@#1\endcsname
\fi
#2}}
\providecommand{\BIBdecl}{\relax}
\BIBdecl

\bibitem{verdecchia2021green}
R.~Verdecchia, P.~Lago, C.~Ebert \emph{et~al.}, ``Green it and green software,'' \emph{IEEE Software}, vol.~38, no.~6, pp. 7--15, 2021.

\bibitem{georgiou2017analyzing}
S.~Georgiou, M.~Kechagia, and D.~Spinellis, ``Analyzing programming languages' energy consumption: An empirical study,'' in \emph{Proceedings of the 21st Pan-Hellenic Conference on Informatics}, 2017, pp. 1--6.

\bibitem{calero_green_2015}
\BIBentryALTinterwordspacing
C.~Calero and M.~Piattini, Eds., \emph{\BIBforeignlanguage{en}{Green in {Software} {Engineering}}}.\hskip 1em plus 0.5em minus 0.4em\relax Cham: Springer International Publishing, 2015. [Online]. Available: \url{https://link.springer.com/10.1007/978-3-319-08581-4}
\BIBentrySTDinterwordspacing

\bibitem{GULDNER2024402}
\BIBentryALTinterwordspacing
A.~Guldner, R.~Bender, C.~Calero \emph{et~al.}, ``Development and evaluation of a reference measurement model for assessing the resource and energy efficiency of software products and components—green software measurement model (gsmm),'' \emph{Future Generation Computer Systems}, vol. 155, pp. 402--418, 2024. [Online]. Available: \url{https://www.sciencedirect.com/science/article/pii/S0167739X24000384}
\BIBentrySTDinterwordspacing

\bibitem{pyRAPL}
\BIBentryALTinterwordspacing
PowerAPI, ``pyrapl: A python library for measuring energy consumption,'' 2023, accessed: 2024-03-05. [Online]. Available: \url{https://github.com/powerapi-ng/pyRAPL/tree/master}
\BIBentrySTDinterwordspacing

\bibitem{noureddine-ie-2022}
A.~Noureddine, ``Powerjoular and joularjx: Multi-platform software power monitoring tools,'' in \emph{18th International Conference on Intelligent Environments (IE2022)}, Biarritz, France, Jun 2022.

\bibitem{MANCEBO2021100558}
\BIBentryALTinterwordspacing
J.~Mancebo, C.~Calero, F.~Garcia \emph{et~al.}, ``Feetings: Framework for energy efficiency testing to improve environmental goal of the software,'' \emph{Sustainable Computing: Informatics and Systems}, vol.~30, p. 100558, 2021. [Online]. Available: \url{https://www.sciencedirect.com/science/article/pii/S2210537921000494}
\BIBentrySTDinterwordspacing

\bibitem{Strubell_Ganesh_McCallum_2020}
\BIBentryALTinterwordspacing
E.~Strubell, A.~Ganesh, and A.~McCallum, ``Energy and policy considerations for modern deep learning research,'' \emph{Proceedings of the AAAI Conference on Artificial Intelligence}, vol.~34, no.~09, pp. 13\,693--13\,696, Apr. 2020. [Online]. Available: \url{https://ojs.aaai.org/index.php/AAAI/article/view/7123}
\BIBentrySTDinterwordspacing

\bibitem{tufano_using_2022}
\BIBentryALTinterwordspacing
R.~Tufano, S.~Masiero, A.~Mastropaolo \emph{et~al.}, ``Using pre-trained models to boost code review automation,'' in \emph{Proceedings of the 44th {International} {Conference} on {Software} {Engineering}}, ser. {ICSE} '22.\hskip 1em plus 0.5em minus 0.4em\relax New York, NY, USA: Association for Computing Machinery, Jul. 2022, pp. 2291--2302. [Online]. Available: \url{https://dl.acm.org/doi/10.1145/3510003.3510621}
\BIBentrySTDinterwordspacing

\bibitem{mastropaolo_studying_2021}
\BIBentryALTinterwordspacing
A.~Mastropaolo, S.~Scalabrino, N.~Cooper \emph{et~al.}, ``\BIBforeignlanguage{en}{Studying the {Usage} of {Text}-{To}-{Text} {Transfer} {Transformer} to {Support} {Code}-{Related} {Tasks}},'' in \emph{\BIBforeignlanguage{en}{2021 {IEEE}/{ACM} 43rd {International} {Conference} on {Software} {Engineering} ({ICSE})}}.\hskip 1em plus 0.5em minus 0.4em\relax Madrid, ES: IEEE, May 2021, pp. 336--347. [Online]. Available: \url{https://ieeexplore.ieee.org/document/9401982/}
\BIBentrySTDinterwordspacing

\bibitem{wang_bridging_2022}
\BIBentryALTinterwordspacing
D.~Wang, Z.~Jia, S.~Li \emph{et~al.}, ``Bridging pre-trained models and downstream tasks for source code understanding,'' in \emph{Proceedings of the 44th {International} {Conference} on {Software} {Engineering}}, ser. {ICSE} '22.\hskip 1em plus 0.5em minus 0.4em\relax New York, NY, USA: Association for Computing Machinery, Jul. 2022, pp. 287--298. [Online]. Available: \url{https://dl.acm.org/doi/10.1145/3510003.3510062}
\BIBentrySTDinterwordspacing

\bibitem{castano_exploring_2023}
\BIBentryALTinterwordspacing
J.~Casta\~{n}o, S.~Mart\'{\i}nez-Fern\'{a}ndez, X.~Franch \emph{et~al.}, ``Exploring the {Carbon} {Footprint} of {Hugging} {Face}'s {ML} {Models}: {A} {Repository} {Mining} {Study},'' in \emph{2023 {ACM}/{IEEE} {International} {Symposium} on {Empirical} {Software} {Engineering} and {Measurement} ({ESEM})}, Oct. 2023, pp. 1--12, arXiv:2305.11164 [cs, stat]. [Online]. Available: \url{http://arxiv.org/abs/2305.11164}
\BIBentrySTDinterwordspacing

\bibitem{SamsiZMLMJBKTG23}
\BIBentryALTinterwordspacing
S.~Samsi, D.~Zhao, J.~McDonald, B.~Li, A.~Michaleas, M.~Jones, W.~Bergeron, J.~Kepner, D.~Tiwari, and V.~Gadepally, ``From words to watts: Benchmarking the energy costs of large language model inference,'' in \emph{{IEEE} High Performance Extreme Computing Conference, {HPEC} 2023, Boston, MA, USA, September 25-29, 2023}.\hskip 1em plus 0.5em minus 0.4em\relax {IEEE}, 2023, pp. 1--9. [Online]. Available: \url{https://doi.org/10.1109/HPEC58863.2023.10363447}
\BIBentrySTDinterwordspacing

\bibitem{lu1codexglue}
S.~Lu, D.~Guo, S.~Ren, J.~Huang, A.~Svyatkovskiy, A.~Blanco, C.~Clement, D.~Drain, D.~Jiang, D.~Tang \emph{et~al.}, ``Codexglue: A machine learning benchmark dataset for code understanding and generation,'' in \emph{Thirty-fifth Conference on Neural Information Processing Systems Datasets and Benchmarks Track (Round 1)}.

\bibitem{dubey2024llama3herdmodels}
\BIBentryALTinterwordspacing
A.~Dubey, A.~Jauhri, A.~Pandey \emph{et~al.}, ``The llama 3 herd of models,'' 2024. [Online]. Available: \url{https://arxiv.org/abs/2407.21783}
\BIBentrySTDinterwordspacing

\bibitem{codecarbon}
\BIBentryALTinterwordspacing
M.~C. Impact, ``Codecarbon: A tool to estimate the carbon emissions of machine learning models,'' 2024, accessed: 2024-03-05. [Online]. Available: \url{https://mlco2.github.io/codecarbon/}
\BIBentrySTDinterwordspacing

\bibitem{HuseinAC25}
\BIBentryALTinterwordspacing
R.~A. Husein, H.~Aburajouh, and C.~Catal, ``Large language models for code completion: {A} systematic literature review,'' \emph{Comput. Stand. Interfaces}, vol.~92, p. 103917, 2025. [Online]. Available: \url{https://doi.org/10.1016/j.csi.2024.103917}
\BIBentrySTDinterwordspacing

\bibitem{shrinkthatfootprint2023}
\BIBentryALTinterwordspacing
S.~T. Footprint, ``Carbon footprint of training gpt-3 and large language models,'' 2023, accessed: 2024-07-22. [Online]. Available: \url{https://shrinkthatfootprint.com/carbon-footprint-of-training-gpt-3-and-large-language-models/}
\BIBentrySTDinterwordspacing

\bibitem{trustbit_llm_benchmarks}
\BIBentryALTinterwordspacing
Trustbit, ``Llm benchmarks,'' 2024, accessed: 2024-07-22. [Online]. Available: \url{https://www.trustbit.tech/en/llm-benchmarks}
\BIBentrySTDinterwordspacing

\bibitem{lmarena_leaderboard}
\BIBentryALTinterwordspacing
L.~Arena, ``Lm arena leaderboard,'' 2024, accessed: 2024-07-22. [Online]. Available: \url{https://lmarena.ai/?leaderboard}
\BIBentrySTDinterwordspacing

\bibitem{oobabooga_benchmark}
\BIBentryALTinterwordspacing
Oobabooga, ``Oobabooga benchmark,'' 2024, accessed: 2024-07-22. [Online]. Available: \url{https://oobabooga.github.io/benchmark.html}
\BIBentrySTDinterwordspacing

\bibitem{strubell2019energypolicyconsiderationsdeep}
\BIBentryALTinterwordspacing
E.~Strubell, A.~Ganesh, and A.~McCallum, ``Energy and policy considerations for deep learning in nlp,'' 2019. [Online]. Available: \url{https://arxiv.org/abs/1906.02243}
\BIBentrySTDinterwordspacing

\bibitem{10.5555/3045118.3045347}
B.~Romera-Paredes and P.~H.~S. Torr, ``An embarrassingly simple approach to zero-shot learning,'' in \emph{Proceedings of the 32nd International Conference on International Conference on Machine Learning - Volume 37}, ser. ICML'15.\hskip 1em plus 0.5em minus 0.4em\relax JMLR.org, 2015, p. 2152–2161.

\bibitem{few-shot}
\BIBentryALTinterwordspacing
R.~L. L.~I. au2, I.~Balažević, E.~Wallace, F.~Petroni, S.~Singh, and S.~Riedel, ``Cutting down on prompts and parameters: Simple few-shot learning with language models,'' 2021. [Online]. Available: \url{https://arxiv.org/abs/2106.13353}
\BIBentrySTDinterwordspacing

\bibitem{LI2024112002}
\BIBentryALTinterwordspacing
X.~Li, S.~Yuan, X.~Gu \emph{et~al.}, ``Few-shot code translation via task-adapted prompt learning,'' \emph{Journal of Systems and Software}, vol. 212, p. 112002, 2024. [Online]. Available: \url{https://www.sciencedirect.com/science/article/pii/S0164121224000451}
\BIBentrySTDinterwordspacing

\bibitem{gholami2021surveyquantizationmethodsefficient}
\BIBentryALTinterwordspacing
A.~Gholami, S.~Kim, Z.~Dong, Z.~Yao, M.~W. Mahoney, and K.~Keutzer, ``A survey of quantization methods for efficient neural network inference,'' 2021. [Online]. Available: \url{https://arxiv.org/abs/2103.13630}
\BIBentrySTDinterwordspacing

\bibitem{faiz2023llmcarbon}
A.~Faiz, S.~Kaneda, R.~Wang, R.~Osi, P.~Sharma, F.~Chen, and L.~Jiang, ``Llmcarbon: Modeling the end-to-end carbon footprint of large language models,'' \emph{arXiv preprint arXiv:2309.14393}, 2023.

\bibitem{10.1145/3540250.3549113}
\BIBentryALTinterwordspacing
C.~Wang, Y.~Yang, C.~Gao, Y.~Peng, H.~Zhang, and M.~R. Lyu, ``No more fine-tuning? an experimental evaluation of prompt tuning in code intelligence,'' in \emph{Proceedings of the 30th ACM Joint European Software Engineering Conference and Symposium on the Foundations of Software Engineering}, ser. ESEC/FSE 2022.\hskip 1em plus 0.5em minus 0.4em\relax New York, NY, USA: Association for Computing Machinery, 2022, p. 382–394. [Online]. Available: \url{https://doi.org/10.1145/3540250.3549113}
\BIBentrySTDinterwordspacing

\bibitem{10.1145/3695988}
\BIBentryALTinterwordspacing
X.~Hou, Y.~Zhao, Y.~Liu, Z.~Yang, K.~Wang, L.~Li, X.~Luo, D.~Lo, J.~Grundy, and H.~Wang, ``Large language models for software engineering: A systematic literature review,'' \emph{ACM Trans. Softw. Eng. Methodol.}, Sep. 2024, just Accepted. [Online]. Available: \url{https://doi.org/10.1145/3695988}
\BIBentrySTDinterwordspacing

\bibitem{10.1145/3661167.3661215}
\BIBentryALTinterwordspacing
C.~Di~Sipio, R.~Rubei, J.~Di~Rocco, D.~Di~Ruscio, and P.~T. Nguyen, ``Automated categorization of pre-trained models in software engineering: A case study with a hugging face dataset,'' in \emph{Proceedings of the 28th International Conference on Evaluation and Assessment in Software Engineering}, ser. EASE '24.\hskip 1em plus 0.5em minus 0.4em\relax New York, NY, USA: Association for Computing Machinery, 2024, p. 351–356. [Online]. Available: \url{https://doi.org/10.1145/3661167.3661215}
\BIBentrySTDinterwordspacing

\bibitem{10.1145/3551349.3559555}
\BIBentryALTinterwordspacing
T.~Ahmed and P.~Devanbu, ``Few-shot training llms for project-specific code-summarization,'' in \emph{Proceedings of the 37th IEEE/ACM International Conference on Automated Software Engineering}, ser. ASE '22.\hskip 1em plus 0.5em minus 0.4em\relax New York, NY, USA: Association for Computing Machinery, 2023. [Online]. Available: \url{https://doi.org/10.1145/3551349.3559555}
\BIBentrySTDinterwordspacing

\bibitem{levenshtein}
G.~Navarro, ``A guided tour to approximate string matching,'' \emph{ACM Computing Surveys}, vol.~33, no.~1, pp. 31--88, 2001.

\bibitem{10.1145/3510003.3510221}
\BIBentryALTinterwordspacing
S.~Georgiou, M.~Kechagia, T.~Sharma, F.~Sarro, and Y.~Zou, ``Green ai: Do deep learning frameworks have different costs?'' in \emph{Proceedings of the 44th International Conference on Software Engineering}, ser. ICSE '22, Springer.\hskip 1em plus 0.5em minus 0.4em\relax New York, NY, USA: Association for Computing Machinery, 2022, p. 1082–1094. [Online]. Available: \url{https://doi.org/10.1145/3510003.3510221}
\BIBentrySTDinterwordspacing

\bibitem{10174114}
S.~Shanbhag and S.~Chimalakonda, ``An exploratory study on energy consumption of dataframe processing libraries,'' in \emph{2023 IEEE/ACM 20th International Conference on Mining Software Repositories (MSR)}.\hskip 1em plus 0.5em minus 0.4em\relax Springer, 2023, pp. 284--295.

\bibitem{bornholt2012model}
J.~Bornholt, T.~Mytkowicz, and K.~S. McKinley, ``The model is not enough: Understanding energy consumption in mobile devices,'' in \emph{2012 IEEE Hot Chips 24 Symposium (HCS)}.\hskip 1em plus 0.5em minus 0.4em\relax IEEE, 2012, pp. 1--3.

\bibitem{jagannadharao_timeshifting_2023}
\BIBentryALTinterwordspacing
A.~Jagannadharao, N.~Beckage, D.~Nafus, and S.~Chamberlin, ``\BIBforeignlanguage{en}{Timeshifting strategies for carbon-efficient long-running large language model training},'' \emph{\BIBforeignlanguage{en}{Innovations in Systems and Software Engineering}}, Dec. 2023. [Online]. Available: \url{https://doi.org/10.1007/s11334-023-00546-x}
\BIBentrySTDinterwordspacing

\bibitem{liu_green_2024}
\BIBentryALTinterwordspacing
V.~Liu and Y.~Yin, ``\BIBforeignlanguage{en}{Green {AI}: exploring carbon footprints, mitigation strategies, and trade offs in large language model training},'' \emph{\BIBforeignlanguage{en}{Discover Artificial Intelligence}}, vol.~4, no.~1, p.~49, Jul. 2024. [Online]. Available: \url{https://doi.org/10.1007/s44163-024-00149-w}
\BIBentrySTDinterwordspacing

\bibitem{rajpurkar2016squad}
P.~Rajpurkar, J.~Zhang, K.~Lopyrev \emph{et~al.}, ``Squad: 100,000+ questions for machine comprehension of text,'' \emph{arXiv preprint arXiv:1606.05250}, 2016.

\bibitem{10.1007/978-3-031-70245-7_12}
V.-A. Cursaru, L.~Duits, J.~Milligan, D.~Ural, B.~R. Sanchez, V.~Stoico, and I.~Malavolta, ``A controlled experiment on the energy efficiency of the source code generated by code llama,'' in \emph{Quality of Information and Communications Technology}, A.~Bertolino, J.~Pascoal~Faria, P.~Lago, and L.~Semini, Eds.\hskip 1em plus 0.5em minus 0.4em\relax Cham: Springer Nature Switzerland, 2024, pp. 161--176.

\bibitem{10.1145/3643916.3644409}
\BIBentryALTinterwordspacing
I.~D. Fagadau, L.~Mariani, D.~Micucci, and O.~Riganelli, ``Analyzing prompt influence on automated method generation: An empirical study with copilot,'' in \emph{Proceedings of the 32nd IEEE/ACM International Conference on Program Comprehension}, ser. ICPC '24.\hskip 1em plus 0.5em minus 0.4em\relax New York, NY, USA: Association for Computing Machinery, 2024, p. 24–34. [Online]. Available: \url{https://doi.org/10.1145/3643916.3644409}
\BIBentrySTDinterwordspacing

\bibitem{10.1145/3411763.3451760}
\BIBentryALTinterwordspacing
L.~Reynolds and K.~McDonell, ``Prompt programming for large language models: Beyond the few-shot paradigm,'' in \emph{Extended Abstracts of the 2021 CHI Conference on Human Factors in Computing Systems}, ser. CHI EA '21.\hskip 1em plus 0.5em minus 0.4em\relax New York, NY, USA: Association for Computing Machinery, 2021. [Online]. Available: \url{https://doi.org/10.1145/3411763.3451760}
\BIBentrySTDinterwordspacing

\bibitem{li2023cctesttestingrepairingcode}
\BIBentryALTinterwordspacing
Z.~Li, C.~Wang, Z.~Liu, H.~Wang, D.~Chen, S.~Wang, and C.~Gao, ``Cctest: Testing and repairing code completion systems,'' 2023. [Online]. Available: \url{https://arxiv.org/abs/2208.08289}
\BIBentrySTDinterwordspacing

\end{thebibliography}


\end{document}